\documentclass[aps,epsf,preprint]{revtex4}

\newcommand{\nslash}{\kern 0.2 em n\kern -0.50em /}

\def\bq{\begin{eqnarray}}
\def\eq{\end{eqnarray}}

\def\roughly#1{\mathrel{\raise.3ex\hbox{$#1$\kern-.75em
\lower1ex\hbox{$\sim$}}}}

\begin{document}

\preprint{\hfill\parbox[b]{0.3\hsize}
{ }}

\def\bra{\langle }
\def\ket{\rangle }

\title{ Neutron single spin asymmetries
from semi-inclusive deep inelastic scattering off 
transversely polarized $^3$He }

\author{Sergio Scopetta\footnote{sergio.scopetta@pg.infn.it}}
\affiliation
{\it 
Dipartimento di Fisica, Universit\`a degli Studi
di Perugia, 
and INFN, sezione di Perugia
\\
via A. Pascoli
06100 Perugia, Italy, 
\\
and Departament de Fisica Te\`orica,
Universitat de Val\`encia, 46100 Burjassot (Val\`encia), Spain }

\begin{abstract}

A study of semi-inclusive deep inelastic scattering
off transversely polarized $^3$He is presented.
The formal expressions
of the
Collins and Sivers contributions to the azimuthal
single spin asymmetry for the production
of leading pions are derived, in impulse approximation, and estimated
in the kinematics of forth-coming experiments
at JLab.
The AV18 interaction has been used for a realistic
description of the nuclear dynamics;
the nucleon structure has been described
by proper parameterizations of data or suitable model calculations.
The initial transverse momentum of the struck quark has
been properly included in the calculation.
The crucial issue of extracting
the neutron information from $^3$He data,
planned to shed some light on the puzzling
experimental scenario arisen from
recent measurements 
for the proton and the deuteron,
is thoroughly discussed. It is found that
a model independent procedure, widely used in inclusive 
deep inelastic scattering to take into account  
the momentum and energy distributions
of the bound nucleons in $^3$He, can be applied also in the 
kinematics of the planned JLab experiments, although
fragmentation functions, not only parton
distributions, are involved. 
The possible role played by final state interactions in the
process under investigation is addressed.

\end{abstract}

\maketitle

\section{Introduction}

The distribution of transverse quark spin, 
the so called transversity distribution,
is one of the least known features of the nucleon partonic structure 
(for a recent review, see, e.g., 
Ref. \cite{bdr}).
In a few years,
some light should be shed on it, thanks to relevant experimental 
efforts 
which are in turn stimulating the theoretical
activity (for recent developments, see Ref. \cite{olmo}).

Semi-inclusive deep inelastic scattering (SIDIS), i.e.
the process $A(e,e'h)X$, with the
detection in the final state
of a produced hadron $h$ in coincidence
with the scattered electron $e'$, is one of the proposed
processes to access the transversity distribution, $h_1$,
and other related observables.
In fact, due to its chiral-odd nature, $h_1$ is not
observable in inclusive deep
inelastic scattering (DIS), while it can appear, multiplied by a
chiral-odd fragmentation function,
in the expression of SIDIS cross-sections.
Since several years it has been known that 
SIDIS off a transversely polarized target
should show azimuthal asymmetries,
the so called ``single spin asymmetries'' (SSAs).
As a matter of facts, it is predicted that
the counting of produced hadrons in a given direction or in the opposite
one, with respect to the reaction plane, 
depends on the orientation of the transverse spin
of a polarized target with respect to the direction
of the unpolarized beam. It can be shown that the SSA
in SIDIS off transverse polarized targets is essentially
due to two different physical mechanisms.
One of them is the Collins mechanism, due to parton final state interactions
in the production of a spinless hadron 
by a transversely polarized quark.
This mechanism produces a term in the SSA 
given essentially by the product of
the transversity distribution and the Collins
fragmentation function. This latter 
quantity is time reversal odd (T-odd) 
and counts
the number density of scalar hadrons
originating from the fragmentation of a
transversely polarized quark
\cite{Collins}.
The other is the Sivers mechanism, producing
a term in the SSA which is given by the product of
the unpolarized fragmentation function with
the Sivers parton distribution
\cite{sivers}, 
naturally related to
the parton Orbital Angular Momentum (OAM),
describing the number density of unpolarized quarks
in a transversely polarized target.

It turns out that, in the experimental asymmetry,
it is technically possible to distinguish   
the different contributions 
of the Collins and Sivers mechanisms
\cite{mu-ta,ko-mu,boer}.

Recently, the first data 
of SIDIS off transverse polarized targets have been published, 
for the proton \cite{hermes} and the deuteron \cite{compass}.
It has been found that, while the Sivers effect
is sizable for the proton, it becomes negligible for the deuteron,
so that apparently the neutron contribution cancels the proton one,
showing a strong flavor dependence of the mechanism.
Very recently, it has been argued that 
these results can be interpreted as an evidence of
a negligible gluon contribution to the OAM
of the partons in the proton \cite{brod}.
In the same paper, it is also stressed that a transversely polarized
$^3$He target, due to its peculiar
spin structure \cite{friar,csps}, could be used to obtain 
relevant information on SSAs of the free neutron, to 
better understand the present experimental scenario.

With the aim at measuring the neutron transversity
and the neutron SSAs,
two experiments of SIDIS off transversely polarized $^3$He targets
have been approved at JLab,
with scientific rate ``A'' , for the detection
of $\pi^-$ (experiment E-06-010 \cite{06010}) and $\pi^+$
(experiment E-06-011 \cite{06011})
in coincidence with the scattered electron, in the valence quark region
($0.19 \le x \le 0.41$), for moderate values of $Q^2$
(1.77 GeV$^2 \le Q^2 \le 2.73$ GeV$^2$).
It has been planned to detect the leading fragmentation pion
which carries $z \simeq 0.5$ of the energy transfer, to favor
current fragmentation.

In this theoretical and experimental scenario, it becomes urgent to confirm
the possibility to use a transversely
polarized $^3$He target to extract
the information for the free neutron.
In the present paper, this issue is thoroughly investigated.
A calculation of SSAs for SIDIS off
transversely polarized $^3$He, described in an impulse 
approximation (IA) framework,
by means of wave functions obtained using a modern nucleon-nucleon
potential, will be illustrated. 
In the calculation, the treatment of the nuclear structure
is realistic, while the nucleon part
relies on model estimates,
being some ingredients of the calculation
experimentally unknown.
As a consequence, the main goal of the paper is not that of producing
realistic predictions but that of establishing to what extent
a transversely polarized $^3$He nucleus behaves as an effective transversely
polarized free neutron.

The paper is structured as follows.
In the second section, the main quantities of interest will be
introduced. In the following one,
the IA expressions for the Collins and Sivers contributions
to the nuclear SSA will be derived and
the calculation will be set-up, in the kinematics of the JLab experiment.
The ingredients to be used will be presented
in the fourth section, together with the numerical results and
their discussion. Eventually, conclusions will be drawn in the last section.

\section{SIDIS and single spin asymmetries}

The quantities to be measured are the Collins, 
$A_{UT}^{Collins}$,
and Sivers, $A_{UT}^{Sivers}$,
contributions to the azimuthal single spin asymmetry, $A_{UT}$
(where $U$ means Unpolarized beam, $T$ means Transversely polarized target
and it is assumed that the produced hadron is spinless,
or that its polarization is not detected) \cite{boer}:
\begin{eqnarray}
A_{UT}(\phi_h,\phi_S) \equiv  
{ 
d \sigma (\phi_h,\phi_S) - d \sigma (\phi_h,\phi_S + \pi)
\over
d \sigma (\phi_h,\phi_S) + d \sigma (\phi_h,\phi_S + \pi)
}
\equiv
{ d^6 \sigma_{UT} \over d^6 \sigma_{UU} }
\label{main}
\end{eqnarray}
where 
the angles are defined according to the ``Trento convention''
\cite{trento}, see Fig. 1.
The cross sections are differential
in $x, y, z, \phi_S, {\bf h_\perp}$, where,
as usual, one has:
\begin{eqnarray}
x & = & {Q^2 \over 2 P \cdot q}~, \\
y & = & {P \cdot q \over P \cdot l} = { \nu \over E_{lab}}~, \\
z & = & {P \cdot h \over P \cdot q} = { E_h^{lab} \over \nu } 
\label{kist}~,
\end{eqnarray}
where $P$, $q$, $h$ are the target momentum, the momentum transfer and
the momentum of the produced hadron, respectively.
A clarification
concerning the notation in SIDIS is in order.
In general,
a quantity which is transverse
in a frame where $P$ and $h$ have no transverse components
(thus ${\bf P_T}={\bf h_{T}}=0$), 
is indicated with a subscript $T$, 
so that $T$ means transverse
with respect to ${\vec h}$, while those with a subscript $\perp$ 
are defined in a frame where $q$ and $P$ have no transverse components
(standard DIS frame), so that $\perp$ means transverse with respect to
$\vec q$ (thus ${\bf q_\perp} = {\bf P_\perp=0}$) \cite{mu-ta}.


Other relevant variables for the description
of this process are $\bf{k_T}$, the transverse momentum of the struck quark
before the interaction, and $\bf{\kappa_T}$, 
the transverse momentum of the struck 
quark after the interaction.
It is found that
$ {\bf{q_T}}  =  - {\bf{h_ \perp}} / z + O(1/Q^2)~.$

The Collins and Sivers asymmetries, according
to the Trento convention \cite{trento},
are defined in terms
of the experimental cross sections:
\bq
A_{UT}^{Collins} & = &
{
\int d \phi_S d \phi_h \sin ( \phi_h + \phi_s )
d^6 \sigma_{UT}
\over
\int d \phi_S d \phi_h  
d^6 \sigma_{UU}
}~,
\label{coll_ex}
\\
A_{UT}^{Sivers} & = &
{
\int d \phi_S d \phi_h \sin ( \phi_h - \phi_S )
d^6 \sigma_{UT}
\over
\int d \phi_S d \phi_h  
d^6 \sigma_{UU}
}~.
\label{siv_ex}
\eq
The cross sections themselves can be
written in terms of ${\bf k_T}$-dependent
distribution and fragmentation functions \cite{mu-ta},
so that also
the Collins and Sivers asymmetries, 
Eqs. (\ref{coll_ex}) and (\ref{siv_ex}),
can be written through the
same functions. In particular,
$A_{UT}^{Collins}$
is found to be:
\begin{eqnarray}
A_{UT}^{Collins} & = &
{
1 - y 
\over
1 - y + y^2/2
}         
| {\bf  S_T} |
\times
\nonumber
\\
& \times & 
{
\sum_q e_q^2 
\int 
d \phi_S
d \phi_h
d^2 {\bf{\kappa_T}}
d^2 {\bf{k_T}}
\delta^2 ( {\bf{k_T}} + {\bf{q_T}}  - {\bf{\kappa_T}} )
{ \bf{\hat{h}} \cdot {\bf{\kappa_T}}  \over M_h}
h_1^q (x, {\bf{k_T}^2} )
H_1^{\perp q,h} (z, (z {\bf{\kappa_T})^2} )
\over
\sum_q e_q^2 
\int 
d \phi_S
d \phi_h d^2 {\bf{\kappa_T}} d^2 {\bf{k_T}}
\delta^2 ( {\bf{k_T}} + {\bf{q_T}}  - {\bf{\kappa_T}} )
f_1^q (x,{\bf{k_T^2}} )
D_1^{q,h}  (z, (z {\bf{\kappa_T})^2} )
}~,
\label{coll}
\end{eqnarray}
where 
$ {\bf S_T} $ is the transverse spin of the target hadron, 
${\bf{\hat h_\perp}} = { \bf{h_ \perp} }/| {\bf {h_ \perp}}|$, 
$M_h$ is the mass of the produced hadron
and
two ${\bf k_T}$ -dependent distributions,
the unpolarized one, 
$f_1^q (x, {\bf{k_T}^2} )$, and the transversity
distribution, $h_1^q (x, {\bf{k_T}^2} )$, appear.
The other two ingredients of Eq. (\ref{coll})
are the ${\bf k_T}$ -dependent unpolarized
fragmentation function
for the production of a scalar hadron,
$D_1^{q,h} (z, ( z {\bf{\kappa_T}} )^2 )$, and the
quantity 
$H_1^{\perp q,h} (z, ( z {\bf{\kappa_T}} )^2 )$,
the so-called Collins fragmentation function,
a T-odd quantity describing 
the number density of scalar hadrons
originating from the fragmentation of a
transversely polarized quark
\cite{Collins}. 

In the same way,
$A_{UT}^{Sivers}$ is found to be:
\begin{eqnarray}
A_{UT}^{Sivers} =
| {\bf  S_T} |
{
\sum_q e_q^2 
\int 
d \phi_S
d \phi_h
d^2 {\bf{\kappa_T}}
d^2 {\bf{k_T}}
\delta^2 ( {\bf{k_T}} + {\bf{q_T}}  - {\bf{\kappa_T}} )
{ \bf{\hat{h}} \cdot {\bf{k_T}}  \over M}
f_{1T}^{\perp q} (x, {\bf{k_T}^2} )
D_1^{q,h} (z, (z {\bf{\kappa_T})^2} )
\over
\sum_q e_q^2 
\int 
d \phi_S
d \phi_h d^2 {\bf{\kappa_T}} d^2 {\bf{k_T}}
\delta^2 ( {\bf{k_T}} + {\bf{q_T}}  - {\bf{\kappa_T}} )
f_1^q (x,{\bf{k_T^2}} )
D_1^{q,h}  (z, (z {\bf{\kappa_T})^2} )
}~,
\label{siv}
\end{eqnarray}
where $M$ is the target mass and, with respect to the
previous asymmetry, Eq. (\ref{coll}), there is a new 
${\bf k_T}$ -dependent
distribution function,
$ f_{1T}^{\perp q} (x, {\bf{k_T}^2} )$.
The latter is a T-odd quantity,
the so called Sivers distribution function,
describing the number density of unpolarized
quarks in a transversely polarized target
\cite{sivers}.

Often, an $x-z$ factorized structure is assumed
for Eqs. (\ref{coll}) and (\ref{siv}),
a procedure which is justified if there is a good experimental
coverage in $| { \bf h_\perp }  |$. 
In fact, in this case, one can integrate 
the numerator and the denominator of Eq. (\ref{main})
over $ {\bf{h_\perp} } = -z {\bf q_T} $,
with proper weights, 
selecting the Collins
and Sivers contributions and getting simplified expressions
for the azimuthal asymmetries \cite{ko-mu,boer}.
Due to the peculiar kinematical conditions of the JLab experiments
\cite{06010,06011},
where $ |{\bf h_\perp}| $ is varying in a narrow range,
such a procedure will not be applied here. The general expressions,
Eqs. (\ref{coll}) and (\ref{siv}), for the Collins and Sivers asymmetries,
will be considered in what follows. 

\section{An Impulse Approximation approach to SIDIS off transversely
polarized $^3$He}

An Impulse Approximation (IA)
approach to perform the first calculation of
the Collins and Sivers Asymmetries
of $^3$He
will be now described.
The idea is that in the SIDIS process 
a single nucleon is interacting with the hard probe and
there are no further
interactions with the recoiling nuclear
system, neither of the jet originating
from the interacting nucleon, 
nor of the emitted hadron. 
The other
crucial assumption of IA is that the internal structure of the bound
nucleon is not different from that of the free one, the nuclear dynamics
determining only its momentum and its binding
energy distributions. 

Now it will be shown that,
in this framework, the Collins and Sivers Asymmetries
of $^3$He, for the production of a hadron $h$,
{\sl in the kinematics of the JLab experiments}, 
are given by the following convolution
expressions:

\begin{eqnarray}
A_{UT,h}^{Collins,^3He}  \simeq 
{
1 - y 
\over
1 - y + y^2/2
} 
| {\bf  S_T} | 
{ N^{Collins,^3He} \over  D^{^3He}}
\label{coll-3}
\end{eqnarray}

and

\begin{eqnarray}
A_{UT,h}^{Sivers,^3He}  \simeq 
| {\bf  S_T} | { N^{Sivers,^3He} \over  D^{^3He}}
\label{siv-3}
\end{eqnarray}

where:

\begin{eqnarray}
N^{Collins,^3He} & = &
\sum_{N=n,p}
\sum_q e_q^2
\int 
d \phi_S
\, d \phi_h
\, d^2 {\bf{\kappa_T}}
\, d^2 {\bf{k_T}}
\, \delta^2 ( {\bf{k_T}} + {\bf{q_T}}  - {\bf{\kappa_T}} )
\nonumber
\\
& \times &
{ \bf{\hat{h}} \cdot {\bf{\kappa_T}}  \over M_h}
H_1^{\perp q,h} (z, (z {\bf{\kappa_T})^2} )
\int_x^{M_3 \over M}
{
d \alpha \over \alpha
}
G^{3,N}_\perp (\alpha) 
h_1^{q,N} 
\left ( { x \over \alpha }, {\bf{k_T}^2} \right )~,
\label{ncol}
\end{eqnarray}

\begin{eqnarray}
D^{^3He} & = &
\sum_{N=n,p}
\sum_q e_q^2
\int d \phi_S
\, d \phi_h
\, d^2 {\bf{\kappa_T}}
\, d^2 {\bf{k_T}}
\, \delta^2 ( {\bf{k_T}} + {\bf{q_T}}  - {\bf{\kappa_T}} )
\nonumber
\\
& \times & D_1^{q,h} 
\left ( z , (z {\bf{\kappa_T}} )^2 \right )
\int_x^{M_3 \over M}
{
d \alpha \over \alpha
}
F^{3,N}(\alpha) 
f_1^{q,N} 
\left ( { x \over \alpha },  {\bf{k_T}^2} \right )~,
\label{dd}
\end{eqnarray}
and
\begin{eqnarray}
N^{Sivers,^3He}  
& =  & 
\sum_{N=n,p}
\sum_q e_q^2
\int 
d \phi_S
\, d \phi_h
\, d^2 {\bf{\kappa_T}}
\, d^2 {\bf{k_T}}
\, \delta^2 ( {\bf{k_T}} + {\bf{q_T}}  - {\bf{\kappa_T}} )
\nonumber
\\
& \times &
{ \bf{\hat{h}} \cdot {\bf{k_T}}  \over M}
D_1^{q,h} (z, (z {\bf{\kappa_T})^2} )
\int_x^{M_3 \over M}
{
d \alpha \over \alpha
}
G^{3,N}_\perp (\alpha) 
f_{1T}^{\perp q,N} \left ( { x \over \alpha}, {\bf{k_T}^2} \right )~.
\label{nsi}
\end{eqnarray}
Now the quantities appearing
in the above equations will be described.
First of all, the variable 
\begin{equation}
\alpha = { \sqrt{2} p^+ \over M} =
{p_0 + p_3 \over M} 
\label{alp}
\end{equation}
represents the light-cone plus-momentum component
of the nucleon $N$, with momentum $p$, to which the struck quark
belongs, divided by the nucleon mass, $M$
\footnote{here and in the following, the
light-cone variables are defined as 
$a^\pm = {1 \over \sqrt{2} } (a_0 \pm a_3)$.}.

The functions $G^{3,N}_\perp (\alpha) $ and $F^{3,N}(\alpha)$ are the 
transverse spin-dependent and the spin-independent 
light-cone momentum
distributions of the nucleon $N$ in $^3$He, respectively, defined as
\begin{equation}
G^{3,N}_\perp (\alpha) = \int dE \int d \vec p 
\, { \sqrt{2} p^+ \over p_0 } \,
P_{\perp}^N(\vec p, E) \delta
\left ( \alpha - { \sqrt{2}p^+ \over M} \right )~,
\label{light_p}
\end{equation}
and
\begin{equation}
F^{3,N}(\alpha) = \int dE \int d \vec p 
\, { \sqrt{2} p^+ \over p_0 } \,
 P^N(\vec p, E) \delta
\left ( \alpha - { \sqrt{2}p^+ \over M} \right )~.
\label{light_u}
\end{equation}
Both these quantities depend on the nuclear structure, being the functions
\begin{equation}
P_{\perp}^N(\vec p, E) = P^N_{ {1 \over 2}{ 1 \over 2},{1 \over 2} }
(\vec p, E)
- P^N_{  -{1 \over 2} -{ 1 \over 2}, {1 \over 2} }
(\vec p, E)
\label{ppar}
\end{equation}
and
\begin{equation}
P^N(\vec p, E) = P^N_{ {1 \over 2}{ 1 \over 2}, {1 \over 2} }
(\vec p, E)
+ P^N_{  -{1 \over 2} -{ 1 \over 2},{1 \over 2} }
(\vec p, E)
~,
\label{pek}
\end{equation}
defined in terms of the components of the spin-dependent
spectral function of the nucleon $N$ in the $^3$He
nucleus, firstly defined in \cite{cps}:
\begin{eqnarray}
 P_{\sigma,
\sigma',{\cal{M}}_x}^{N} ({\vec{p},E}) & = & \sum\nolimits\limits_{{f}_{(A-1)}}
{}~_{N}\langle{\vec{p},\sigma_x;\psi
}_{f_{(A-1)}} |{\psi }_{J{\cal{M}}_x}\rangle~
\nonumber
\\
& &
\langle{\psi
}_{J{\cal{M}}_x}|{\psi }_{f_{(A-1)}};\vec{p},\sigma_x '\rangle _{N}~
\delta (E-{E}_{f_{(A-1)}}+{E}_{A})~,
\label{eq9}
\medskip
\end{eqnarray}
where $|{\psi}_{J{\cal{M}}_x}\rangle$ 
is the ground state of the target nucleus polarized along
the $x$ axis, $|{\psi }_{f_{(A-1)}}\rangle$  an eigenstate of the $(A-1)$ 
nucleon
system interacting with the same two-body potential of the target nucleus,
and $|\vec{p},\sigma_x\rangle_N$  the plane wave state
for the nucleon $N$ with the spin projection
along the $x$-axis equal to $\sigma_x$.
From Eqs. (\ref{alp})-(\ref{light_u}) it is clear that, in the
extreme non-relativistic
(NR) limit, i.e. with nucleons at rest,
$\alpha =1 $ and no nuclear effects are found.
The size of the leading nuclear effects discussed here is of
the order $|\vec p|/M$.

One should notice that, in this framework, 
the probability to find a transversely polarized
nucleon in a transversely polarized nucleus is the same
of that of finding a longitudinally polarized nucleon
in a longitudinally polarized nucleus, since the only possible
treatment of the nuclear system is NR. It turns out therefore
that the transverse and longitudinal spin dependent light-cone momentum
distributions are equal and that the results obtained for the latter
can be used also for the first quantity, the one of interest here.
In other words, the overlaps $_{N}\langle{\vec{p},\sigma_x;\psi
}_{f_{(A-1)}} |{\psi }_{J{\cal{M}}_x}\rangle$
and $_{N}\langle{\vec{p},\sigma_z;\psi
}_{f_{(A-1)}} |{\psi }_{J{\cal{M}}_z}\rangle$
turn out to be equal in a NR framework (for the
identity of the helicity and transversity distributions
in a NR framework see, e.g., Ref \cite{ja_er,bac}).

The main lines of the formal derivation of Eqs. 
(\ref{coll-3}) 
and (\ref{siv-3}) will be now discussed.
It is a generalization of
the standard procedure for obtaining the nuclear parton
distributions of $^3$He in DIS. In particular, the derivation
of the expression in the numerator of Eq. (\ref{coll-3})
and (\ref{siv-3}) generalizes
the one of the $^3$He helicity distribution, to
be found in \cite{csps,sauer,guz}, while 
that of the denominator generalizes
the I.A. derivation
of the $^3$He unpolarized parton distribution, which can be found, e.g.,
in Ref. \cite{ss}. With respect to the cited papers,
the difference here is the presence of the fragmentation
functions and of the ${\bf k_T}$-dependence in the distribution functions
in the SIDIS cross-sections.
Let us start from the general expressions Eqs. (\ref{coll})
and (\ref{siv}).
In order to evaluate them for a generic target $A$, one
has to estimate therefore three ${\bf k_T}$-dependent parton
distributions,$f_1^{q,A}, h_1^{q,A}, f_{1T}^{\perp q, A }$, 
and two ${\bf k_T}$-dependent fragmentation
functions, $D_1^{q,A,h}, H_1^{\perp q,A,h}$.
Let us discuss how to obtain the nuclear effects in I.A.
for one of the  ${\bf k_T}$-dependent parton distributions
and one of the ${\bf k_T}$-dependent fragmentation functions.

As an example, let us consider the parton distribution
$f_1^{q,A}$, whose definition reads (see, e.g., \cite{bdr}):
\begin{eqnarray}
f_1^{q,A}( x, {\bf k_T^2} ) = 
\int
{ d \xi^- d {\bf \xi_T} \over 2 (2 \pi)^3 }
e^{i ( xP^+ \xi^- - {\bf k_T} \cdot {\bf \xi_T} ) }
\langle P | \bar \Psi(0,0,0) 
\, \gamma^+ \,
\Psi( 0, \xi^- , {\bf \xi_T} ) | P  \rangle~, 
\label{fdef}
\end{eqnarray}
where the light-cone target states $| P \rangle$ 
are normalized according to
\bq
\bra P'  | P \ket = (2 \pi)^3 2 P^+ \delta(P'^+ - P^+)
\delta^2(\vec P_\perp - \vec P_\perp')~.
\label{lcns}
\eq
By substituting in Eq. (\ref{fdef}) the standard expansion
of the quark field $\Psi$, writing for simplicity
the quark degrees of freedom only, one gets:
\begin{eqnarray}
f_1^{q,A}(x,{\bf k_T^2}) & = & \int
{ d \xi^- d {\bf \xi_T} \over 2 (2 \pi)^3}
e^{i(xP^+\xi^- - 
{\bf k_T} \cdot {\bf \xi_T}  
)}
\nonumber
\\
& &
\int { d k'^+ {\bf d k'_T} \over 2 k'^+ (2 \pi)^3}
\langle P | \hat O
e^{-i(k'^+\xi^- - 
{\bf k'_T} \cdot {\bf \xi_T}
)}
| P \rangle
\nonumber
\\
& = & {1 \over 2 P^+}
\int { d k'^+ {\bf d k'_T} \over 2 k'^+ (2 \pi)^3}
\, \langle P | 
\hat O \,
| P \rangle
\delta \left( x - { k'^+ \over P^+} \right )
\delta^2 \left( {\bf k'_T} - {\bf k_T}  \right )
\label{ffield}
\end{eqnarray} 
where $ \hat O $ is the quark operator:
\bq
\hat O = \sum_r 
\bar u_r(k'^+, {\bf k'_T}) b_r^\dag (k'^+, {\bf k'_T})
\, \gamma^+ \,
b_r(k'^+, {\bf k'_T}) u_r(k'^+, {\bf k'_T})
\eq
and the
creation
and annihilation operators,
$b^\dag(k)$ and
$b(k)$, obey the commutation relation 
\bq
\{ b(k'),b^\dag(k)\} = (2 \pi)^3 2 k^+ \delta(k'^+ - k^+)
\delta^2({\bf k_T} -  {\bf k_T'})~.
\eq

Let us think now to a nuclear target with $A$ nucleons.
Following a standard procedure, in the above equation
two complete sets of states, corresponding
to the interacting nucleon in an IA scenario
and to the recoiling nuclear system, are 
properly inserted 
to the left and right-hand sides of the quark operator:

\begin{eqnarray}
f_1^{q,A}(x,{\bf k_T^2})  & = & 
{1 \over 2 P^+} 
\int { d k'^+ {\bf d k'_T} \over 2 k'^+ (2 \pi)^3 }
\langle P | 
 \,\, \sum_{\vec P_R',S_R',\vec p',s'} 
{ \{ | \vec P_R' S_R' \ket | \vec p' s' \ket \}} 
\{ \bra \vec P_R' S_R' |  \bra \vec p' s'| \} 
\nonumber
\\
& \hat O & 
 \sum_{\vec P_R,S_R,\vec p,s} 
\{ | \vec P_R S_R \ket | \vec p s \ket \} 
\{ \bra \vec P_R S_R |  \bra \vec p s| \} 
\,\,\,\,
| P \rangle
\, \delta \left( x - { k'^+ \over P^+} \right )
\delta^2 \left( {\bf k'_T} - {\bf k_T}  \right )~.
\label{ia}
\end{eqnarray}
Since later 
the nuclear matrix elements have to be evaluated by means
of NR wave functions,
the inserted states are normalized in a NR manner:
\bq
\bra \vec P'  | \vec P \ket = (2 \pi)^3 
\delta^3(\vec P' - \vec P)~, 
\label{nrns}
\eq
so that one has $|P \ket = \sqrt{2 P_0} |\vec P \ket$ (cf. Eq.
(\ref{lcns}) )\cite{muld}. 
Due to this normalization and since, using IA and
translational invariance,
\begin{eqnarray}
{\{ \bra \vec P_R S_R |  \bra \vec p s| \} 
| \vec P S \ket} = {\bra \vec P_R S_R, \vec p s  
| \vec P S \ket } (2 \pi)^3 \delta^3 (\vec P - \vec P_R - \vec p)
\delta_{S,S_R\,s}~,
\end{eqnarray}
Eq. (\ref{ia}) can be rewritten:

\begin{eqnarray}
f_1^{q,A}(x,{\bf k_T^2})  & = & 
{P_0 \over  P^+} \int d \vec p
\,\,  \sum_{S_R,s} 
\langle \vec P  
 | \vec P_R S_R, \vec p s \ket   
 \bra \vec P_R S_R, \vec p s| \vec P \rangle 
\nonumber
\\
& \times & {1 \over 2 p_0} 
\int { d k'^+ {\bf d k'_T} \over 2 k'^+ (2 \pi)^3 } \,\,
\bra p s| \hat O | p s \ket  
\delta \left( x - { k'^+ \over P^+} \right )
\delta^2 \left( {\bf k'_T} - {\bf k_T}  \right )~.
\label{temp}
\end{eqnarray}
In the nuclear rest frame (RF), where $M_A = P_0 = \sqrt{2} P^+$,
introducing the Bjorken variable $ x_B = {Q^2 \over 2 M \nu} = 
{M_A \over M} x$, and using the definition of $\alpha$, Eq. (\ref{alp}),
one of the $\delta$ functions in the above equation can be written: 
\bq
\delta \left( x - { k'^+ \over P^+} \right ) =
{1 \over \alpha} \delta \left( { x \over \alpha}
 - { k'^+ \over \alpha P^+} \right ) =
{1 \over \alpha}
{M_A \over M} \delta \left( {x_B \over \alpha} - { k'^+ \over p^+} \right )~,
\eq
so that Eq. (\ref{temp}), using 
the definition of the unpolarized nuclear spectral function,
Eq. (\ref{pek}), and the expression of $f_1^{q,A}(x,{\bf k_T^2})$
for a generic $A$ target, Eq. (\ref{ffield}), becomes:
\bq
f_1^{q,A}(x,{\bf k_T^2}) 
= {M_A \over M}
\sum_N \int dE \int d \vec p \,\, P^N(\vec p,E) \,\,
{ \sqrt{2} p^+ \over p_0} {1 \over \alpha}
f_1^{q,N} \left( {x_B \over \alpha}, {\bf k_T^2} \right )~.
\label{fin}
\end{eqnarray}

Eq. (\ref{fin}) gives
the ${\bf k_T}$-dependent nuclear parton distribution
in a convolution-like form, in terms of the nuclear spectral function and
of the ${\bf k_T}$-dependent parton distribution
for the internal, moving nucleon $N$. 
Similar expressions can be obtained for the other two
nuclear ${\bf k_T}$-dependent parton
distributions, $h_1^{q,A}$ and $f_{1T}^{\perp q, A }$, appearing in
Eqs. (\ref{coll}) and (\ref{siv}). 

Concerning the nuclear effects on the fragmentation functions,
let us consider the standard unpolarized ${\bf k_\perp}$-dependent
one, $D_1^{q,A,h}(z, (z{\bf \kappa_T})^2 )$.
Since the hadron $h$ is produced by the fragmentation
of a quark belonging to a nucleon with four-momentum $p$
in the nuclear medium, in an IA scenario, where
the nucleon structure is not modified by the nuclear medium,
$D_1^{q,A,h}$ is given by the corresponding quantity for the nucleon,
with $z= {p \cdot h \over p \cdot q}$, i.e.:

\bq
D_1^{q,A,h}(z, z^2{\bf \kappa_T^2} ) 
= 
D_1^{q,N,h}\left ( {p \cdot h \over p \cdot q}, 
\left( {p \cdot h \over p \cdot q} \right )^2{\bf \kappa_T^2} 
\right )~.
\label{d1qa}
\eq
A similar expression can be obtained for the other 
nuclear ${\bf k_T}$-dependent fragmentation function,
$H_1^{\perp q,A,h}$, appearing in
Eq. (\ref{coll}). 

Due to the general structure of Eqs. (\ref{fin}) and (\ref{d1qa}),
it is therefore possible to write the final formulae
for the Sivers and Collins contributions to the single spin asymmetry
of $^3He$, in I.A., in the nuclear R.F.:

\begin{eqnarray}
A_{UT}^{Collins} = 
{
1 - y 
\over
1 - y + y^2/2
} 
| {\bf  S_T} | 
{ N^{Collins} \over D}~,
\label{coll_ia}
\end{eqnarray}

\begin{eqnarray}
A_{UT}^{Sivers} =
| {\bf  S_T} |  
{ N^{Sivers} \over D}~,
\label{siv_ia}
\end{eqnarray}

with:

\begin{eqnarray}
N^{Collins}
& = & 
\sum_N
\int dE \int d \vec p 
\int d \alpha \, \delta \left ( \alpha -  {\sqrt{2} p^+ \over M}
\right )  
P_\perp^N(\vec p, E) {\sqrt{2} p^+ \over p_0} {1 \over \alpha}
\nonumber
\\
& \times & 
\sum_q e_q^2 \int d^2 {\bf{\kappa_T}}
d^2 {\bf{k_T}}
\delta^2 ( {\bf{k_T}} + {\bf{q_T}}  - {\bf{\kappa_T}} )
{ \bf{\hat{h}} \cdot {\bf{\kappa_T}}  \over M_h}
\nonumber
\\
& \times & 
h_1^{q,N} \left( { x_B \over \alpha}, {\bf{k_T}^2} \right )
H_1^{\perp q,N,h}  \left ( {p \cdot h \over p \cdot q}, 
\left ( {p \cdot h \over p \cdot q} \right )^2{\bf \kappa_T^2} 
\right )~,
\label{n_col}
\end{eqnarray}

\begin{eqnarray}
D & = &
\sum_N
\int dE \int d \vec p 
\int d \alpha \, \delta \left ( \alpha -  {\sqrt{2} p^+ \over M}
\right )  
P^N(\vec p, E) {\sqrt{2} p^+ \over p_0} {1 \over \alpha}
\nonumber
\\
& \times & 
\sum_q e_q^2 \int d^2 {\bf{\kappa_T}}
d^2 {\bf{k_T}}
\delta^2 ( {\bf{k_T}} + {\bf{q_T}}  - {\bf{\kappa_T}} )
\nonumber
\\
& \times & 
f_1^{q,N} \left(  {x_B \over \alpha}, {\bf{k_T}^2} \right )
D_1^{q,N,h} \left ( {p \cdot h \over p \cdot q}, 
\left ( {p \cdot h \over p \cdot q} \right )^2{\bf \kappa_T^2} 
\right )~,
\label{den}
\end{eqnarray}

\begin{eqnarray}
N^{Sivers} & = &
\sum_N
\int dE \int d \vec p 
\int d \alpha \, \delta \left ( \alpha -  {\sqrt{2} p^+ \over M}
\right )  
P_\perp^N(\vec p, E) {\sqrt{2} p^+ \over p_0} {1 \over \alpha}
\nonumber
\\
& \times & 
\sum_q e_q^2 \int d^2 {\bf{\kappa_T}}
d^2 {\bf{k_T}}
\delta^2 ( {\bf{k_T}} + {\bf{q_T}}  - {\bf{\kappa_T}} )
{ \bf{\hat{h}} \cdot {\bf{k_T}}  \over M}
\nonumber
\\
& \times & 
f_{1T}^{\perp q,N} 
\left(  {x_B \over \alpha}, {\bf{k_T}^2} \right )
D_1^{q,N,h} \left ( {p \cdot h \over p \cdot q}, 
\left ( {p \cdot h \over p \cdot q} \right )^2{\bf \kappa_T^2} 
\right )~.
\label{n_siv}
\end{eqnarray}

I note in passing that the term  
${\sqrt{2} p^+ \over p_0}$,
appearing in Eqs. (\ref{n_col})-
(\ref{n_siv}), is
the so called flux-factor
\cite{ku,fs}, clearly emerging
when one goes from field theoretical definitions
in light-cone quantization to a NR framework, 
in order to use
the available NR nuclear wave functions.

Eqs. (\ref{n_col})-(\ref{n_siv}) show that the Fermi motion
and binding effects predicted by I.A.
affect the longitudinal
momentum in the parton distribution of the bound nucleon,
and the longitudinal and transverse momentum
in the fragmentation function of a struck quark belonging
to a bound nucleon. The evaluation of Eqs. (\ref{coll_ia})
and (\ref{siv_ia}) is therefore rather involved and the extraction
of the quantities of interest from experimental data looks
quite cumbersome. Nevertheless, in the kinematics
of the JLab experiments E-06-010 and E-06-011,
everything becomes simpler.
As a matter of facts, the chosen
kinematics helps in two different ways.
First of all, to favor pions from  current fragmentation, 
the variable $z$ has been chosen to be around the value of 0.5,
which means that, 
in the kinematics of the proposed experiment,
only ultra-relativistic pions 
with an energy $E_h \simeq p_h \simeq 2.4$ GeV are detected.
Secondly, the pions are detected in a narrow cone around the direction
of the momentum transfer $\vec q$.
The maximum value of $\theta_{hq}$, the angle
between the directions of the
virtual photon and of the emitted pion is around 12$^o$
\cite{06010}. Therefore, also the angle between $\vec p$ and $\vec q$,
$\theta_{pq}$,
and the one between $\vec p$ and $\vec h$, 
$\theta_{ph}$, are very similar.
This makes possible to observe that, in DIS kinematics 
($\nu \simeq | \vec q|$):
\bq
{p \cdot h \over p \cdot q} & \simeq &
{ E_h \, ( p_0 - p \cos \theta_{ph})
\over 
\nu \, ( p_0 - p \cos \theta_{pq}) }
= {E_h \over \nu} \left \{ 1 - {p \over p_0}
( \cos \theta_{ph} - \cos \theta_{pq} ) +  
O \left [ \left ( p \over p_0 \right )^2 \right ] \right \}
\nonumber
\\
& = & 
{ E_h \over \nu } \left \{ 1 -  {p \over p_0} 
\sin \left ( 
{ \theta_{ph} + \theta_{pq} \over 2 } 
\right )
\sin \left ( 
{ \theta_{pq} - \theta_{ph} \over 2 } 
\right )
+  O \left[
\left ( p \over p_0 \right )^2 \right ] \right \}
\simeq {E_h \over \nu } = z~,
\label{nofrag}
\eq
the last identity holding due to the little difference
between $\theta_{pq}$ and 
$\theta_{ph}$. As a matter of facts, the 
leading nuclear effects discussed here,
affecting the nuclear parton distributions,
are of the order  $ O \left ( p \over p_0 \right )$, and therefore
the $p$-dependence in the argument
of the fragmentation functions,
given by $p/ p_0$ multiplied by a vanishing coefficient,
yields subleading effects which
can be disregarded. 

Eventually, inserting
Eq. (\ref{nofrag}) in the Eqs (\ref{n_col}) - (\ref{n_siv}),
Eqs. (\ref{coll_ia})
and (\ref{siv_ia}) reduce to Eqs. (\ref{coll-3}) 
and (\ref{siv-3}), which are therefore
the expressions to be evaluated in the kinematics of
the JLab experiments.

Applying the proposed formalism, 
the SSAs for $^3$He can be estimated by calculating Eqs. (\ref{coll-3}) 
and (\ref{siv-3}), i.e. by modelling the functions
$G_\perp^{3,N}(\alpha)$
and $F^{3,N}(\alpha)$,
dependent on
the nuclear structure, and the five functions 
$f_1^{q,N}, h_1^{q,N}, f_{1T}^{\perp q, N },
D_1^{q,h}, H_1^{\perp q,h}$,
depending on the nucleon structure.
The nuclear part can be properly treated
by using a realistic nuclear polarized spectral
function, Eq. (\ref{eq9}), describing the Fermi motion
and binding effects, which allows one to evaluate
the light-cone momentum distributions, 
$G_\perp^{3,N}(\alpha)$
and $F^{3,N}(\alpha)$, according to Eqs. (\ref{light_p}) - 
(\ref{pek}).
The nucleon part can be evaluated by using fits of experimental
data, whenever they are available, or results of model calculations.
The ingredients used in the present analysis
will be discussed in the next section.

Now the main issue of extracting the neutron information
from the nuclear asymmetries will be addressed.
To this aim, in \cite{06010,06011}, it has been
proposed to use
a method,
the one suggested and justified, for the first time 
for DIS processes, in 
\cite{csps}. In that paper,
it was shown that in polarized DIS nuclear effects described
by IA, i.e., Fermi motion and binding effects, are safely taken care of
by considering the nucleons with an
{\sl effective polarization}, safely predictable with realistic
nuclear wave functions \cite{friar}.
In the present case,
the idea leads to approximate the experimental asymmetry for $^3$He,
$A_3^{exp}$, by the following expression:
\begin{equation}
A_3^{exp,i}  \simeq  2 f_p p_p A_{\vec p}^i + f_n p_n A_{\vec n}^i~,
\label{mod2}
\end{equation}
where $f_{p(n)}$
is the proton (neutron) 
``dilution factor'', the index $i$ 
means Collins or Sivers,
$A_{{\vec p}(\vec n)}^i$
is the proton (neutron) asymmetry
and the effective nucleon polarizations are:
\begin{eqnarray}
p_p & = & 
\int 
dE \int d \vec p \, P_\perp^p(\vec p E)
=-0.028{\pm} 0.004~,
\label{polp} \\
p_n & = & 
\int 
dE \int d \vec p \, P_\perp^n(\vec p E)
=0.86 {\pm} 0.02~,
\label{pol}
\end{eqnarray}
as obtained by averaging many few-body calculations,
available at the time of publication of Ref. \cite{friar}
\footnote{Since that time, other
calculations have been performed. For example,
in Table II of Ref. \cite{kpsv}, some values of effective polarizations
obtained by considering also the Coulomb repulsion between the protons
in $^3$He, not included in the analysis of Ref. \cite{friar}, are
reported.
}.
One should notice that, were polarized $^3$He a perfect 
effective neutron target with all the nucleons in $S$ wave,
one would find $p_n=1$ and $p_p=0$.

Eventually, the dilution factors are:
\begin{eqnarray}
f_{p(n)}(x,z)=
{\sum_q e_q^2
f_1^{q,{p(n)}} 
\left ( x \right )
D_1^{q,h} \left ( z  \right )
\over
\sum_{N=p,n}
\sum_q e_q^2
f_1^{q,N} 
( x )
D_1^{q,h} 
\left ( z \right )
}~.
\label{dilut}
\end{eqnarray}
While the approximation Eq. (\ref{mod2}), theoretically
proposed and justified
in \cite{csps}, has been widely used 
in the treatment of DIS data,
the possibility of using it also in SIDIS deserves a careful analysis,
which will be performed in the next section.
Since now it is worth to stress 
that, if Eq. (\ref{mod2}) were a good approximation of reality,
it would be possible to use it
to extract the neutron Asymmetry according to the
following formula, suggested in \cite{csps}:
\begin{equation}
A^i_n \simeq {1 \over p_n f_n} \left ( A^{exp,i}_3 - 2 p_p f_p
A^{exp,i}_p \right )~.
\label{extr}
\end{equation}
If one uses in Eq. (\ref{extr}) a 
realistic calculation to simulate the experimental data
$A^{exp,i}_3$, and the model used for the proton 
in the calculation
to simulate
the experimental data for $A^{exp,i}_p$, if nuclear effects were
safely taken care of by Eq. (\ref{mod2}), one should be able
to extract, according to Eq. (\ref{extr}), the neutron asymmetry
used as an input for the calculation. 
In the same way, the neutron information would be obtained safely
from nuclear and proton data using Eq. (\ref{extr}), 
if the main nuclear structure effects
were properly described by Eq. (\ref{mod2})

\section{Results and discussion}

As stated in the previous section, in order to calculate
the SSAs of $^3$He, Eqs. (\ref{coll-3}) 
and (\ref{siv-3}), one nuclear structure ingredient,
the polarized spectral
function, Eq. (\ref{eq9}), providing the light-cone momentum
distributions $G_\perp^{3,N}(\alpha)$
and $F^{3,N}(\alpha)$, and five nucleon structure ingredients,
the functions $f_1^{q,N}, h_1^{q,N}, f_{1T}^{\perp q, N },
D_1^{q,h}, H_1^{\perp q,h}$, have to be modelled.

In the following, the quantities used in the calculation
are presented and discussed.

Concerning the nuclear part, the spin-dependent polarized spectral
function,  Eq. (\ref{eq9}), corresponding to the Argonne V18 interaction
(AV18, \cite{av18}), has been evaluated.
This has been done along the line of Ref. \cite{kpsv}, using the
overlaps calculated in \cite{pssk} by means of a wave function
obtained by the method of Ref. \cite{pisa}.
The procedure gives, for the proton and the neutron in $^3$He, the effective
polarizations $p_p=-0.024$ and 
$p_n=0.878$, respectively, in agreement with the analysis
of Ref. \cite{friar}, Eqs. (\ref{polp}) and
(\ref{pol}),
and only slightly different from the predictions
of the AV14 interaction 
($p_p=-0.026$, $p_n=0.873$ \cite{kpsv}), as expected.

Concerning the nucleon, to perform the calculation, 
the following five ingredients have been
used:

\begin{enumerate}

\item{}
for the unpolarized ${\bf k_T}$-dependent parton
distribution, $f_1^{q,N}(x,{\bf k_T}^2 )$,
use has been made of the usual 
Gaussian factorization:
\begin{equation}
f_1^{q,N}(x,{\bf k_T}^2)= f_1^{q,N}(x) G({\bf k_T}^2)~,
\label{ansatz}
\end{equation}
with \cite{ans}:
\begin{equation}
 G({\bf k_T}^2) = { 1 \over \pi < {\bf k_T}^2 > }
e^{ - {\bf k_T}^2  / < {\bf k_T}^2 >}   ~,
\end{equation}
where
$ < {\bf k_T}^2 > = $ 0.25 GeV$^{2}$,
and for the standard parton distribution $f_1^{q,N}(x)$
use has been made of the
parameterization of Ref. \cite{gluu};

\item{}
for the transversity distribution,
$h_1^{q,N}$, still experimentally
poorly known, use has been made of 
the same Gaussian factorization above, with the ansatz 
$h_1^q(x) = g_1^q(x)$, i.e.,
the transversity distribution has been taken to be
equal to the standard helicity distribution.
This gives certainly the correct order of magnitude.
In particular, the parameterization of Ref. \cite{glup}
has been used;

\item{}
for the Sivers function,
$f_{1T}^{\perp q} (x, {\bf{k_T}^2} )$ in Eq. (\ref{siv-3}),
use has been made of a fit of the few available data,
proposed in Ref. \cite{ans};

\item{}
for the unpolarized fragmentation function
$D_1^{q,h}(z,(z{\bf \kappa_T})^2 )$, different models are used for
evaluating the Sivers and Collins Asymmetries.
For evaluating the Sivers one, 
use has been made of the Gaussian factorization:
\begin{equation}
D_1^{q,h}(z,(z{\bf \kappa_T})^2)= D_1^{q}(z) \, G [ (z{\bf \kappa_T})^2 ]~,
\label{ansatz1}
\end{equation}
with \cite{ans}:
\begin{equation}
 G({\bf p_T}^2) = { 1 \over \pi < {\bf p_T}^2 > }
 e^{- {\bf p_T}^2 / < {\bf p_T}^2 >   }~,
\end{equation}
where
$ < {\bf p_T}^2 > = $ 0.2 GeV$^{2}$, and using
for the standard fragmentation function
$D_1^{q}(z)$ the
parameterization in Ref. \cite{kret}.
For evaluating the Collins Asymmetry, use has been made
of the model calculation of 
$D_1^{q,h}(z,(z{\bf \kappa_T})^2 )$
discussed in Ref. \cite{amr}, 
in particular, the one performed
considering pseudoscalar pion-quark coupling, 
Eq. (6) of that paper.
The latter choice has been done for consistency reasons, since
the same scenario has been used to model
the Collins fragmentation function in evaluating
the Collins Asymmetry, as explained in the following point;

\item{}
for the basically unknown Collins fragmentation function,
$H_1^{\perp q}(z, (z {\bf \kappa_T})^2 )$, appearing
in Eq. (\ref{coll-3}), use has been made
of the model calculation of Ref. \cite{amr}, 
in particular, of the contribution arising from pion loops,
considering pseudoscalar pion-quark coupling, 
Eqs. (8) -- (12) of that paper.

\end{enumerate}

To fix the ideas, results will be shown for the production
of $\pi^-$, although everything can be immediately extended to
$\pi^+$ production.

The results for the SSAs of $^3$He are
not very illuminating by themselves, being
the calculation strongly dependent on the models used
as an input for the nucleon transverse spin structure, i.e. on the
models chosen for the basically unknown
$h_1$ distribution,
Sivers distribution and Collins
fragmentation function. In presenting the results,
it is better therefore to emphasize the 
relevance of nuclear effects in the extraction of the
neutron information. A convenient scheme to illustrate them
is the following. One could assume that the full calculation,
i.e. the evaluation of Eqs. (\ref{coll-3}) and (\ref{siv-3}),
using the ingredients listed above, represents a set
of experimental data. The neutron information is then
extracted from them by using different models of the $^3$He
spin structure. The obtained result is then compared to
the neutron SSA used as an input in the calculation.
The closer the obtained curves are, the better
the proposed extraction procedure works.
This would allow to proper take into account Fermi
motion and binding effects in obtaining the neutron
information, without performing difficult deconvolutions
of the Eqs. (\ref{coll-3}) and (\ref{siv-3}), once the data
are available.
Of course this procedure does not take into account
effects beyond Fermi motion and binding, i.e. beyond
IA.
This scheme is used to present the results in Figs. 2 and 3.
The free neutron asymmetry used as a model in the
calculation, given by a full line, is compared with two curves.
One is the quantity:
\begin{equation}
\bar A^i_n \simeq {1 \over f_n} A^{exp,i}_3~,
\label{extr-1}
\end{equation}
where  $A^{exp,i}_3$ is the result of the full calculation, 
simulating data, and $f_n$ is the neutron dilution
factor, Eq. (\ref{dilut}).
Eq. (\ref{extr-1}) is the relation one would expect to hold
between the $^3$He and the neutron SSAs if there were no nuclear effects,
i.e. $^3$He were a system of free nuleons in a pure $S$ wave.
In fact it can be obtained from Eq. (\ref{extr}) by 
imposing $p_n=1$ and $p_p=0$.
$\bar A^i_n $ is given by the dotted curve in the figures.
The third curve, the dashed one, is obtained by using
Eq. (\ref{extr}), i.e. thinking to $^3$He as a nucleus
where the effects of its complicated
spin structure, leading to a depolarization
of the bound neutron, together with the ones of
Fermi motion and binding, can be easily taken care
of, according to Eq. (\ref{mod2}),
by parameterizing the nucleon effective polarizations
of $^3$He. For the latter
quantities, in Eq. (\ref{extr}) 
the values given in Eqs. (\ref{polp}) and (\ref{pol})
from Ref. \cite{friar},
not the ones obtained by the AV18 interaction
used in the full calculation, have been chosen.
This is done to take into account the possible error
due to the choice of the nucleon-nucleon potential.

In Fig. 2 it is shown the result of the procedure for the Collins
Asymmetry, corresponding to the production of
$\pi^-$, as a function of $x_B$ and for various
values of $z$. The values $Q^2 = 2.2$ GeV$^2$
and $\theta_{qh}= 10^o$,
central values in the kinematics of the
TJNAF experiment \cite{06010}, have been chosen
for the calculation.
In the latter experiment, data will be taken
for $ 0.13 < x_B  <0.41 $, $0.46 <z < 0.59$
and $E_{lab} = 6$ GeV.
It is clear from the figure that the difference 
between the full and dotted curve,
showing the amount of nuclear effects, is sizable,
being around 10 - 15 \% for any $x$ and $z$ values
in the range relevant for the planned experiment,
while the difference between the dashed
and full curves reduces drastically
to a few percent, showing that the extraction
scheme Eq. (\ref{extr}) takes safely into account
the spin structure of $^3$He, together with Fermi
motion and binding effects.
One should remember that 
in the proposed experiments only the region
$0.13 < x_B < 0.41$ will
be actually explored.

In Fig. 3 the same is shown for the Sivers Asymmetry,
again in the kinematics of
the proposed TJNAF experiment.
In this region, for the Sivers Asymmetry,
the same conclusions 
discussed previously for the Collins one hold.
I reiterate that one should not take the absolute
size of the Asymmetries too seriously,
being the obtained neutron Collins and Sivers asymmetry 
strongly dependent on the models chosen for the unknown distribution
functions. One should instead consider the difference between
the curves which are shown, a model independent
feature which is the most relevant outcome of the present
investigation.

From Figs 2 and 3
it is clear therefore that Eq. (\ref{extr}) will be a valuable tool
for the data analysis of
the experiments \cite{06010} and \cite{06011}.

In principle, effects beyond IA could be relevant
in the process under scrutiny.
Among them, shadowing effects and final state interactions
(FSI) could play a role.
As a matter of facts, some doubts on the possibility
of using safely the extraction procedure Eq. (\ref{extr}) due
to shadowing effects have arisen \cite{strik}.
One should remember anyway that these effects are supposed
to be active for values of Bjorken $x$ smaller
than the ones explored by the planned experiments. 
From this point of view, problems should not arise.
Concerning FSI, 
the assumption that even the outgoing hadron, which can be slow,
does not interact further, 
makes the use of the IA in SIDIS more questionable with respect to the
standard inclusive DIS case.
Moreover, since the planned values of
$Q^2$ are smaller than 3 GeV$^2$, in principle the effects
of FSI cannot be
ignored. One should notice anyway that, in the planned experiments,
the energy of the emitted pions is high, being 
$0.46 <z < 0.59$, chosen to give
$E^\pi \simeq$ 2.4 GeV.
The latter observation should make rather safe the use of IA results for
the analysys of the JLab experiments.
Moreover, 
in the numerator of the asymmetry, Eq. (\ref{main}), the effects of
any spin-independent interaction should cancel out, in particular
for a spinless hadron as the pion.
In any case, the analysis of FSI
for this complicated process deserves a dedicated analysis.
This work is in progress, together with the inclusion in the scheme
of other more realistic models of the nucleon structure, able to
predict reasonable figures for the experiments.

\section{Conclusions}

The recent measurement of the single spin asymmetry
for the proton and the deuteron exhibited the importance
of safely accessing the same quantity for the neutron.
As for any polarized neutron observable, $^3$He,
due to its peculiar spin structure, is the natural
target. Two experiments, aimed at measuring azimuthal asymmetries
in the production of $\pi^\pm$  from transversely
polarized $^3$He, have been approved at JLab.
It is planned to obtain the neutron information from $^3$He data,
using a procedure that performs well in the inclusive DIS case,
being able to take care of several crucial aspects of nuclear dynamics,
such as the spin structure of $^3$He 
and the momentum and energy distributions
of bound nucleons.
Here, the problem whether or not that procedure
can be extended to the present SIDIS scenario,
involving therefore fragmentation functions and not only parton
distributions, has been thoroughly analyzed,
in impulse approximation.
The general formulae including the nuclear structure effects
are derived, and evaluated in the specific JLab kinematics,
using the AV18 interaction
to describe the nuclear structure and 
parameterizations of data or suitable model calculations
to describe the nucleon part. 
The initial transverse momentum of the struck quark has
been properly included in the calculation.
It is found that,
in the kinematics of the proposed experiments,
Fermi motion and binding effects can be safely taken
care of by a simple extraction procedure, where the only nuclear
structure ingredients are the nucleon effective polarizations, 
quantities known from precise few body calculations
in a rather model independent way.
The possible role of final state interactions,
and the evaluation of the quantities of interest
by means of realistic models of the nucleon, to get  
figures for the planned experiments, have been addressed
and are under investigation.

\acknowledgments
\vskip 2mm
I am grateful to S. Noguera and V. Vento for fruitful discussions,
and to the Department of Theoretical Physics of the Valencia
University, where this study was mainly done,
for warm hospitality.
This work is supported in part by the INFN-CICYT agreement
and by the Generalitat Valenciana under the contract 
AINV06/118.

\newpage

\section*{Figure Captions}

\vspace{1em}\noindent
{\bf Fig. 1}:
Definition of the azimuthal angles for the process 
$A \uparrow (e,e'h)X$, 
in the rest frame of the target $A$, 
according to the ``Trento convention''
(after Ref. \cite{trento}).

\vspace{1em}\noindent
{\bf Fig. 2}:
The model neutron Collins asymmetry for
the production of $\pi^-$
(full), and the one extracted
from the full calculation according to Eq. (\ref{extr})
(dashed) and
Eq. (\ref{extr-1}) (dotted), 
for $z = 0.3, 0.45, 0.6$, in panels (a), (b), (c), respectively,
for $Q^2= 2.2$ GeV$^2$.

\vspace{1em}\noindent
{\bf Fig. 3}:
The same as in Fig. 2, but for the Sivers asymmetry.

\newpage

\begin{figure}[ht] 
\includegraphics{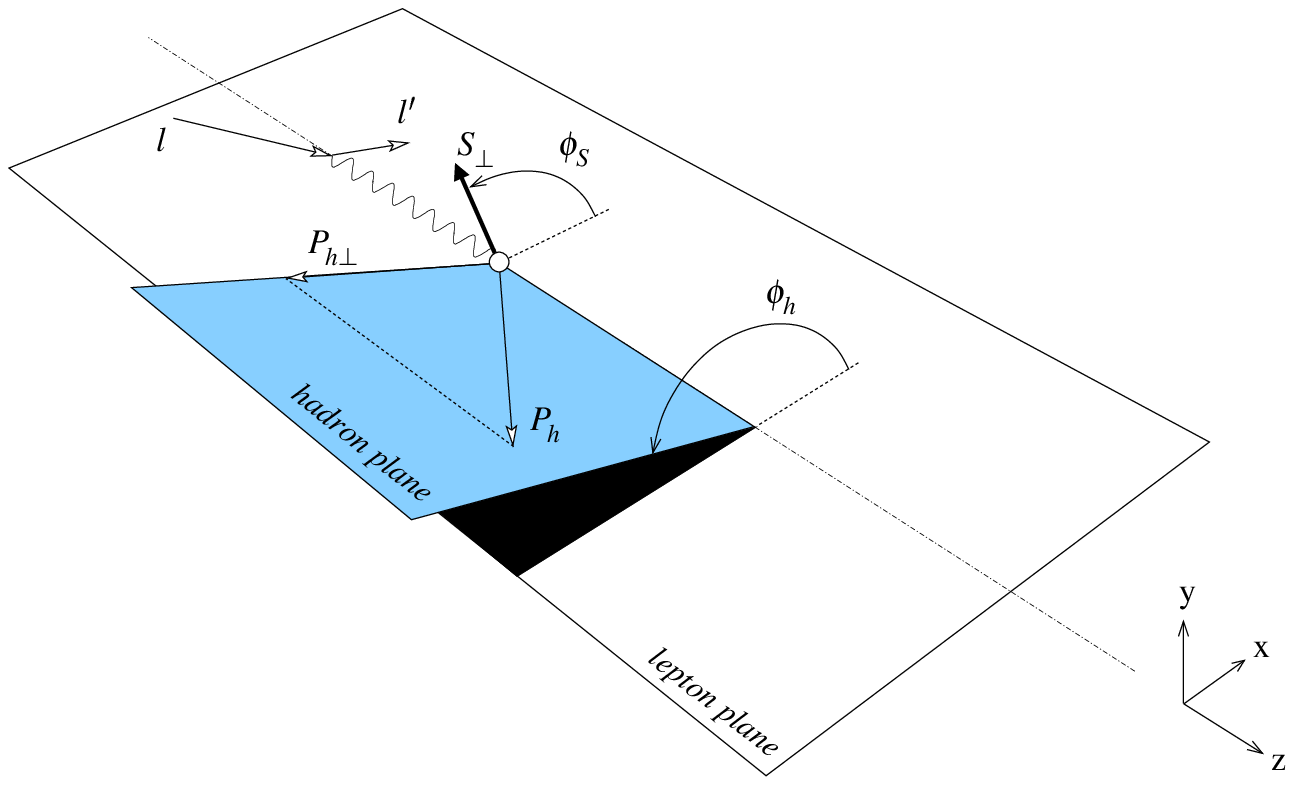}
\vspace{12.0cm}
\caption{}
\end{figure}

\newpage

\begin{figure}[h]
\includegraphics{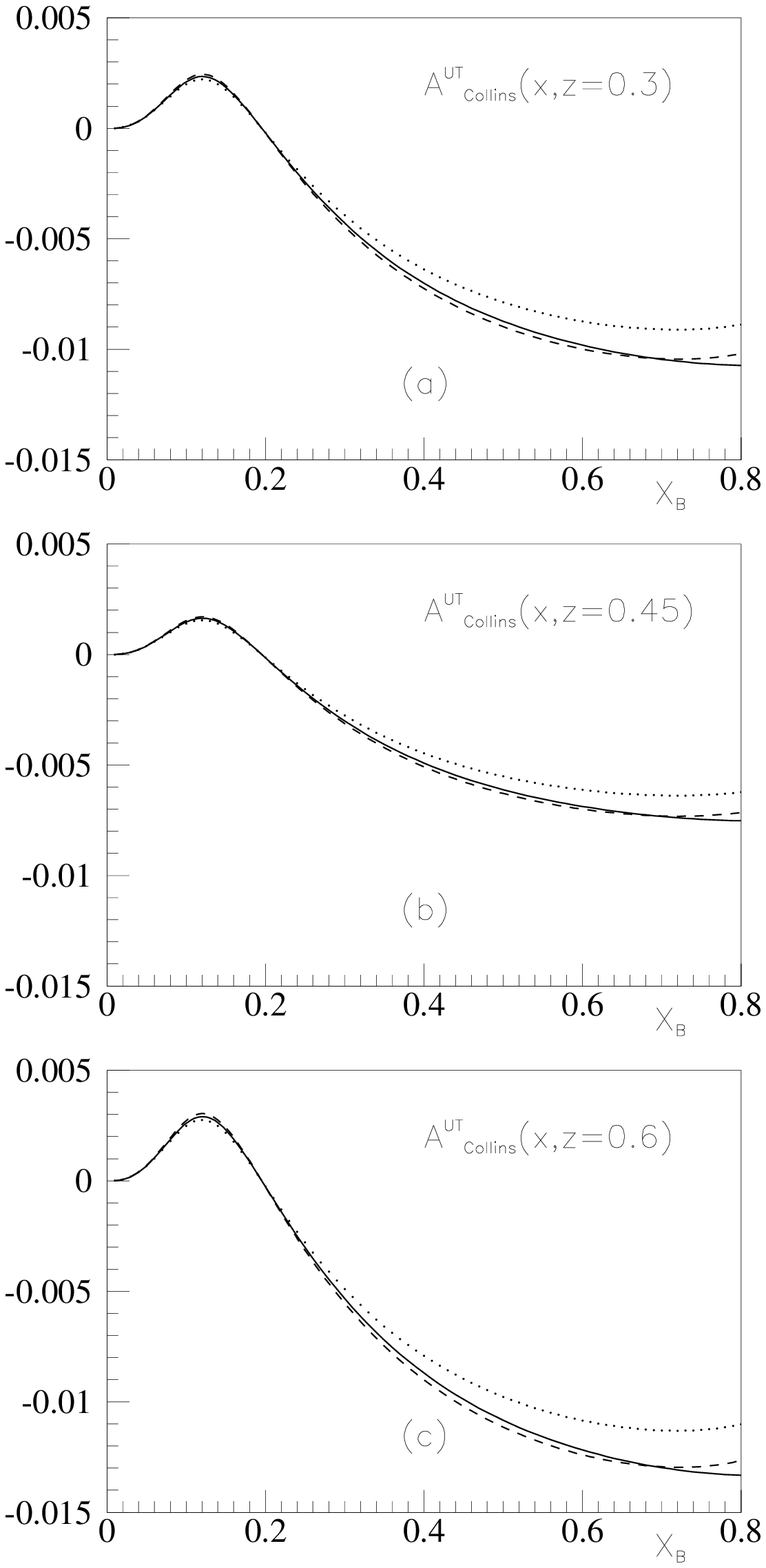}
\vspace{12.0cm}
\caption{}
\end{figure}

\newpage

\begin{figure}[h]
\includegraphics{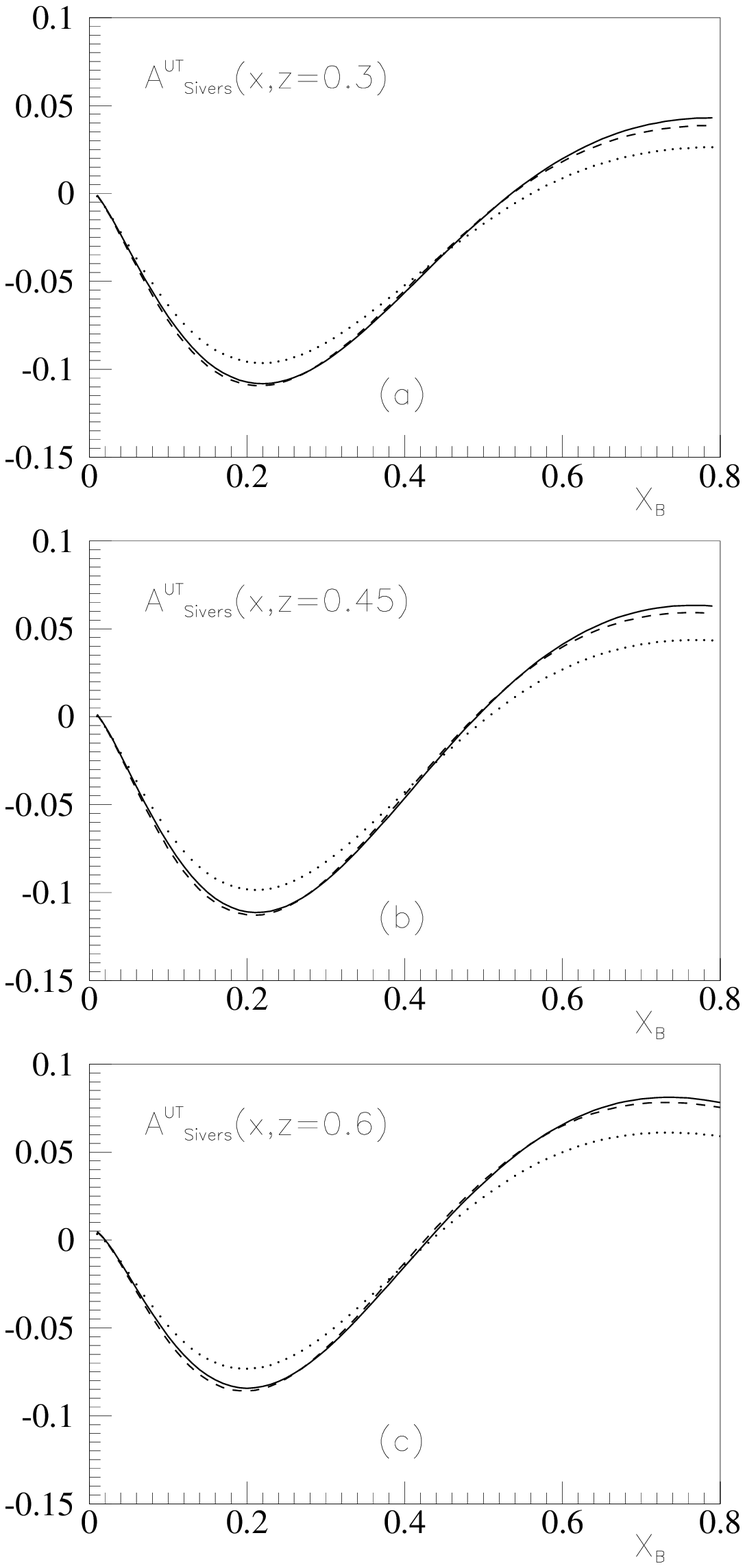}
\vspace{12.0cm}
\caption{}
\end{figure}

\end{document}